\documentclass[]{aa}  
\usepackage{natbib}
\bibpunct{(}{)}{;}{a}{}{,} 
\usepackage{hyperref}
\usepackage{ulem}
\usepackage{xcolor}
\usepackage{graphicx}
\usepackage{txfonts}
\usepackage{orcidlink}
\usepackage{bm}

\begin{document}

\title{Statistical analysis of solar energetic particle rise times using Earth and Mars observations and constraints on particle transport parameters}
\titlerunning{SEP rise time at Earth and Mars}

\author{Yihang Cao\inst{1}\orcidlink{0009-0008-9762-0703}, Jingnan Guo\inst{1, $*$}\orcidlink{0000-0002-8707-076X}, Yuming Wang\inst{1}\orcidlink{0000-0002-8887-3919}, Zhuxuan Zou\inst{1}\orcidlink{0009-0008-9920-9600}, Yongjie Zhang\inst{2}\orcidlink{0000-0002-5746-8064}, Cunhui Li\inst{3}\orcidlink{0000-0003-2689-9387}}

\institute{Deep space Exploration Laboratory/School of Earth and Space Sciences, University of Science and Technology of China, Hefei, PR China \and Institute of Modern Physics, Chinese Academy of Sciences, Lanzhou, China \and National Key Laboratory of Space Environment Interation with Matters, Lanzhou Institute of Physics}

\abstract
{The propagation of solar energetic particles (SEPs) in interplanetary space is modulated by solar wind turbulence, which significantly influences particle diffusion and energy evolution through scattering processes. However, traditional analyses based on absolute flux measurements face inherent difficulties in disentangling the effects of particle acceleration at the source from subsequent transport processes. In contrast, the temporal features of SEP, such as the onset time and the peak time, are less affected by these uncertainties and therefore are more suitable for exploring transport process of SEP.}
{This study aims to establish a statistical relationship between the rise time of SEP events at different energies using multi‑satellite observations at Earth and Mars. The results are used to further invert the turbulence parameters affecting the transport of SEPs based on a parallel diffusion model, thereby revealing the characteristics and evolutionary trend of turbulence at different radial distances.} 
{Using data from SOHO/ERNE and Tianwen-1/MEPA between November 2020 and March 2025, we selected 75 SEP events at 1 AU and 58 events near Mars. For each energy range of each event, the onset time was determined by a linear fitting method, and the peak time was extracted using a sliding median filter combined with Savitzky-Golay smoothing methods. The difference between the onset time and the peak time subsequently determines the SEP rise time. Together with the theoretical relation of rise time and its dependence on energy derived from the pure diffusion equation, we compared the statistical behavior of rise time between Earth and Mars.}
{Despite uncertainties introduced by selection criteria of SEP events, the rise time of SEP events, $\Delta t$ in minute, follows a clear power‑law relation with energy $(\Delta t\propto E^\eta)$, with $\eta = -0.368\pm0.058$ averaged for SEP events near Earth and $\eta = -0.302\pm0.031$ averaged for events near Mars. 
The flatter power‑law shape at Mars suggests that the dependence of rise time with energy becomes less pronounced with increasing radial distance from the Sun. 
Based on these empirical relations, we constrain the rigidity dependence the parallel mean free path of SEPs within the parallel diffusion model.
Using the simple assumption of a pure diffusion model, our results indicate that turbulence scattering of energetic particles at Mars approaches a rigidity‑independent regime, reflecting an evolutionary trend of turbulence toward a dissipation‑dominated state from Earth to Mars.}
{}

\keywords{Sun: particle emission -- Sun: activity}

\maketitle
\nolinenumbers

\section{Introduction} \label{sec:1}
Solar Energetic Particle (SEP) events are energetic particle populations originating from solar eruption activities, consisting of protons, electrons, and heavy ions, with energies typically ranging from suprathermal (a few keV) to relativistic (a few GeV) \citep{desai2016large}. Our current understanding of the spatial distribution and temporal evolution of SEPs has benefited significantly from recent multi-satellite observations \citep{GUO2024}. However, considerable uncertainties in flux measurements arise from instrument cross-calibration discrepancies, particle anisotropy, and contamination of low-energy channels by high-energy particle deposits \citep[e.g.,][]{lario2013longitudinal,li2020identification}, which pose further challenges for quantitative studies of SEP transport. On the other hand, temporal parameters such as the onset time and the peak time of flux are less affected by these factors, making them more robust diagnostic parameters for multi-satellite SEP event analysis \citep{meyer1956solar, krucker2000two}. Therefore, investigating how rise times (time duration between onset and peak times) of SEP events vary with radial distance is crucial for understanding transport mechanisms of particles in the interplanetary space.

The analysis of rise time of SEP events is grounded in the interpretation of the particle time-intensity profile, which includes both the source injection and transport information. The theoretical study of SEP transport dates back to \citet{meyer1956solar}, who interpreted the time profile of the ground-level enhancement (GLE) events using the diffusion equation. This framework was later extended by \citet{parker1965passage} into the classic transport equation, which comprehensively describes particle motion under diffusion, adiabatic cooling, and magnetic focusing in the interplanetary magnetic field. This equation is typically solved numerically with assumed particle injection functions for comparison with observations due to its complexity \citep[e.g.,][]{heras1992influence, Heras1995, droge1994transport, qin2006effect, zhang2009propagation, hu2017modeling, whitman2023review}.

From an observational perspective, the SEP time-intensity profile is often described as a product of a power-law and an exponential function \citep{farwa2025superposed}. Consequently, the Weibull distribution proposed by \citet{wilks2011statistical} has been widely used to fit the observed time-intensity profiles. Subsequent studies, such as \citet{kahler2017characterizing}, employed a modified Weibull function for the fitting which has a general form as:
\begin{align}
    F(t)=\left|\frac{a}{b}\right|\left(\frac{t}{b}\right)^{a-1}\cdot\text{exp}\left[-\left(\frac{t}{b}\right)^a\right].\label{eqn1.1: Modified_Weibull}
\end{align}
Here, $F$ represents the intensity of SEPs, $t$ is the time from event onset, the positive parameter $a$ defines the profile shape, and $b$ is the scaling parameter that stretches or compresses the basic shape of the event along the $t$-axis.

\citet{laurenza2016weibull} and \cite{pallocchia2017weibull} suggested that the Weibull distribution can serve as an asymptotic solution to the diffusion-loss equation, while \citet{chiappetta2021proton} proposed a connection to stochastic acceleration mechanisms in turbulence. However, the shape parameter $a$ and the scaling parameter $b$ lack clear physical correspondence in the Weibull distribution, limiting its utility for explaining specific physical processes.

Some studies have focused on simplified scenarios to isolate individual physical effects. \citet{krimigis1965interplanetary} noted that under diffusion-dominated conditions, the evolution of the SEP intensity simplifies to the solution of a pure diffusion equation. Building on this, \citet{wang2022quantitative} further incorporated a radial dependence of the diffusion coefficient and derived an analytical expression for the rise time ($\Delta t$) from onset to peak.

The scattering experienced by SEPs during propagation originates primarily from the solar wind turbulence. The solar wind, approximated as a collisionless plasma, serves as a natural laboratory for studying magnetohydrodynamic and kinetic-scale turbulence \citep{bruno2013solar}. The power spectral density (PSD) of magnetic turbulence in wavenumber space typically exhibits segmented power-law features, divided into three characteristic ranges: the energy-containing range, the inertial range, and the dissipation range \citep{bruno2013solar, duan2018angular}.

In the energy-containing range, energy is injected by large-scale motions. In the inertial range, energy cascades to smaller scales, and the spectral index of magnetic PSD is a key identifier of the turbulence state. For instance, the classic Kolmogorov theory of isotropic turbulence predicts a spectral index of $-5/3$, while the Iroshnikov–Kraichnan (IK) model for magnetohydrodynamic turbulence predicts an index of $-3/2$ \citep{kolmogorov1941local, iroshnikov1964turbulence, kraichnan1965inertial, kolmogorov1991local}. 
As turbulence develops to the ion kinetic scales and further to the electron kinetic scales, it enters the dissipation range \citep{lotz2023radial}. In this region, wave behavior becomes dispersive and energy is converted into thermal energy via dissipation or wave-particle interactions, causing a steepening of the PSD power-law slope. Between the ion and electron kinetic scales, the magnetic PSD typically shows a spectral index ranging from approximately $-2$ to $-4$, a region sometimes referred to as the ``secondary inertial range'' or ``electron inertial range'' \citep{horbury2008anisotropic, bruno2013solar}. The precise structure of PSD at scales near or even smaller than the electron kinetic scale remains a frontier research topic.

This multi-scale turbulent structure directly determines the scattering efficiency experienced by charged particles propagating through it. Within the inertial range of magnetic turbulence, where the PSD follows a power law $P(k)\propto k^{-q}$, quasi-linear theory and the resonance scattering condition yield the relation $\alpha = 2-q$, which links the turbulence spectral index $q$ to the rigidity dependence index $\alpha$ of the mean free path \citep[][see details in Sect. \ref{sec:results.2}]{bieber1994proton}. Therefore, inferring $\alpha$ from observed SEP propagation characteristics, such as comparing the SEP rise time information at Earth and Mars, can also provide important independent constraints on the nature of solar wind turbulence and its influence on SEP transport at different heliospheric distances.

Currently, Solar and Heliospheric Observatory (SOHO) at Earth and Tianwen‑1 at Mars provide a unique multi‑satellite observation opportunity to systematically investigate the variation of SEP rise times at different distances. In this paper, we analyze SEP rise times based on observations from these two spacecraft between November 2020 and March 2025. In Sect. \ref{sec:data} and Sect. \ref{sec:events}, we depict the datasets and event selection criteria, respectively. In Sect. \ref{sec:results}, we establish the correlation between rise time and proton energy, and invert turbulence parameter based on a diffusion transport model. Section \ref{sec:conclusion} provides a summary and discussion of our findings.

\section{Data description}\label{sec:data}
SOHO was launched in December 2, 1995 and has been operating near the Sun–Earth L1 Lagrange point since early 1996 \citep{torsti1995energetic, domingo1995soho}. For this study, we utilize proton flux data from the Energetic and Relativistic Nuclei and Electron experiment (ERNE) onboard SOHO, specifically from its 9 proton energy channels covering 13 to 100 MeV. It should be noted that the SOHO spacecraft executes a roll maneuver approximately every three months, switching its orientation between $0^\circ$ and $180^\circ$. This can affect the instrument's alignment relative to the interplanetary magnetic field and therefore the primary angles of particle coming into the telescope, but its impact on the statistical results of rise times is considered to be minor.

The Wind spacecraft, launched on November 1994, has been operating in a halo orbit around the Sun-Earth L1 Lagrange point since its insertion in 1996. The Magnetic Field Investigation (MFI) onboard Wind provides vector magnetic field measurements with high precision \citep{lepping1995wind}. In this study, we use the 10 Hz magnetic field data for turbulence analysis.

Tianwen-1 was launched on July 23, 2020 and entered Mars orbit in August 2021. The orbiter of the mission follows a highly elliptical orbit with an inclination angle of $86.9^\circ$, a periapsis altitude of $\sim265$ km, a peroapsis altitude of $\sim12,000$ km, and an orbital period of $7.8$ hr. The Mars Energetic Particle Analyzer (MEPA) onboard Tianwen-1 has been continuously monitoring energetic particles after entering Mars orbit \citep{tang2020calibration, li2021design}. With a time resolution of 4 seconds, MEPA provides proton flux measurements across 15 energy channels spanning 2 to 100 MeV. In addition to particle measurements, the Mars Orbiter Magnetometer (MOMAG) onboard Tianwen-1 measures the magnetic field from the solar wind to the magnetic pileup region around Mars. The calibration procedures, in-flight performance, and comparison with MAVEN magnetic field data can be found in the references \citep{yu2023tianwen, zou2023flight, wang2023mars}. In this study, we use MOMAG data sampled at a high cadence of 32 Hz to characterize the turbulence properties during SEP events.

In addition to particle and magnetic field measurements, we also incorporate radiation dose data from the Trace Gas Orbiter (TGO) \citep{semkova2018charged}. A key objective of TGO is the long term monitoring of the Martian orbital radiation environment. 
We found that TGO's dose rate measurements during SEP events provide a very good assessment of the rise time for high-energy protons reaching Mars, offering an independent constraint on the relationship between rise time and energy beyond the range of MEPA's direct particle measurements.

\section{Events analyses} \label{sec:events}
\subsection{Event selection}
We examine SEPs observed by SOHO/ERNE and Tianwen-1/MEPA between November 2020 and March 2025. We use both datasets with a time resolution of one minute. Event selection is based on the following criteria: (1) All selected events must exhibit a clear rise phase, requiring that the 4-hour interval preceding the peak intensity be free of any significant enhancements attributable to other SEP events. However, events that overlap the decay tail of a prior event are retained if the decay flux is below $10\%$ of the current peak intensity and does not distort the rise profile. (2) Only events with sufficiently high peak intensity were included to guarantee robust statistics. The thresholds are set at $1~ \text{MeV}^{-1}\text{s}^{-1}\text{sr}^{-1}\text{cm}^{-2}$ in the 2-2.6 MeV channel of Tianwen-1/MEPA, and $10^{-2}~ \text{MeV}^{-1}\text{s}^{-1}\text{sr}^{-1}\text{cm}^{-2}$ in the 13-16 MeV channel of SOHO/ERNE. As a result, 58 events are selected at Mars and 75 events are selected at Earth for the subsequent analysis.

\subsection{Rise time determination}
Calculating rise time requires determining both the onset time and the peak time for each energy channel of each SEP event. Two widely used methods for identifying the onset time are the Poisson-Cusum method \citep{huttunen2005proton} and the linear fitting method \citep{xie2016energy}. The Poisson‑Cusum method pre‑defines a background interval and a sliding window length, identifying the first data point where values consecutively exceed a Cusum statistical threshold as the onset time. The latter first calculates the mean and standard deviation of a pre-event background interval, and $3\sigma$ is adopted as the background deviation threshold. A linear fit is then performed on the flux data during the rapid rise phase, and the time at which the fitted line intersects the background deviation threshold is defined as the onset time.

\begin{figure*}[ht!]
\begin{minipage}[t]{1.0\linewidth}
\centering
\includegraphics[width=1.0\hsize]{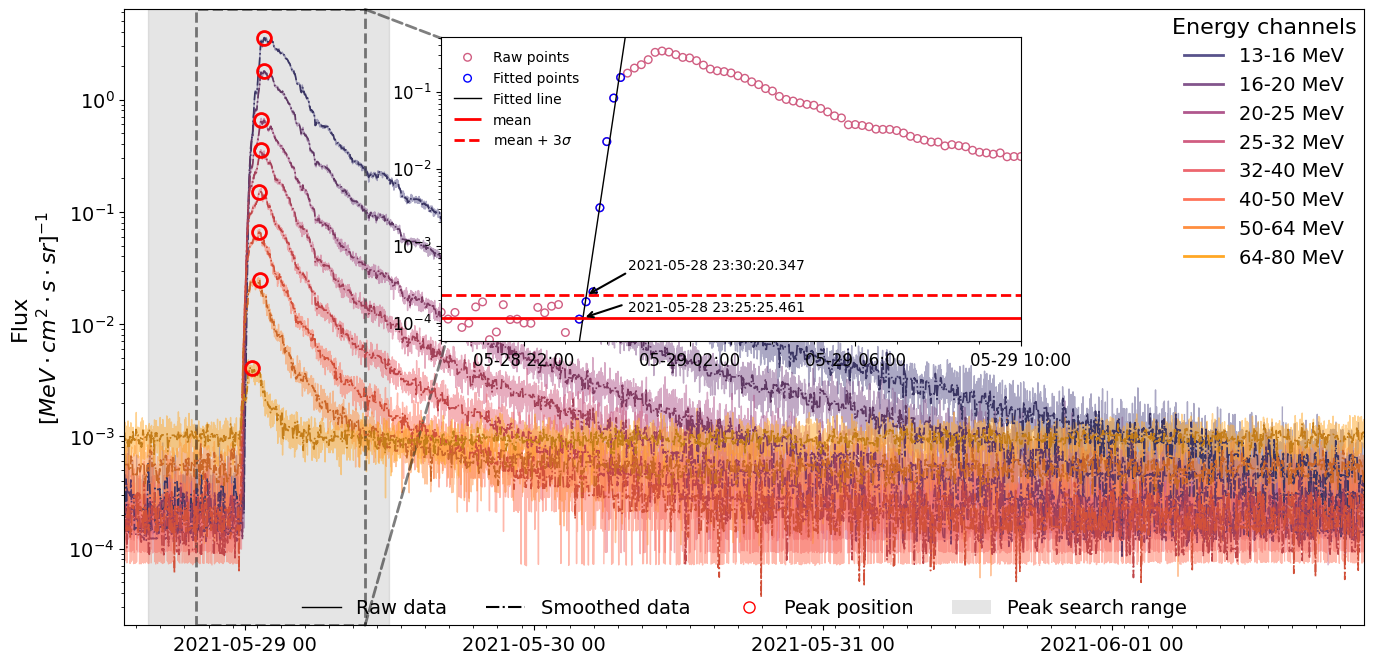} 
\caption{Illustration of the methodology for determining the onset and peak times of an SEP event observed by SOHO/ERNE on May 28, 2021. The main panel shows time-intensity profiles across eight energy ranges (13–100 MeV), where the raw data are plotted as solid lines and the smoothed data as dash-dotted lines. The gray shaded region denotes the search window for peak identification, with identified peaks marked by red circles. The inset zooms into the dashed region, displaying the onset determination for 25-32 MeV range using the linear fitting method. The raw data points are shown in pink circles, and the blue circles represent the data points used for the linear fit, with the black solid line showing the fit result. The horizontal red lines indicate the background level (solid line for mean and dashed line for 3$\sigma$). The intersection of the black fitted line with the red background line defines the onset time.}
\label{fig:1}
\end{minipage}
\end{figure*}

In this study, we employ the linear fitting method because it yields onset times with better precision and is less sensitive to inherent temporal resolution of the data, thus providing more precise results for most of the events studied. An example illustrating this procedure is shown in Fig.\ref{fig:1}. To reduce the influence of short term fluctuations on the linear fitting method, the onset determination is performed using flux data averaged over 10-minute intervals, downsampled from the original 1-minute resolution. Despite this averaging, the fitted onset time retains precision at a higher level because it is derived from the intersection of the fitted line with the background level.

The most straightforward method for determining the peak time is to select the time corresponding to the maximum flux within a chosen interval. However, for high resolution data, random fluctuations due to less statistics can introduce significant uncertainty. Therefore, we adopted an optimized algorithm for peak identification: First, we applied a sliding median filter to the original flux time series to remove outlier data points, and then smoothed the data using a Savitzky–Golay filter, which reduces high frequency noise while preserving waveform characteristics \citep{savitzky1964smoothing, schafer2011savitzky}. Within a defined search window for the peak, we set a prominence threshold equal to 0.3 times the variance of the smoothed data within that window, which helped distinguish the true main peak from minor fluctuations. Finally, we only accepted a candidate peak if the time interval to any adjacent peak was at least 5 minutes and its full width at half maximum exceeded 3 minutes, ensuring the identification of a physically reliable peak time. Despite the previous smoothing procedure used to identify the peak, the final peak time is identified from the 1-minute resolution data to capture the exact maximum.

\begin{figure*}[ht!]
\begin{minipage}[t]{1.0\linewidth}
\centering
\includegraphics[width=1.0\hsize]{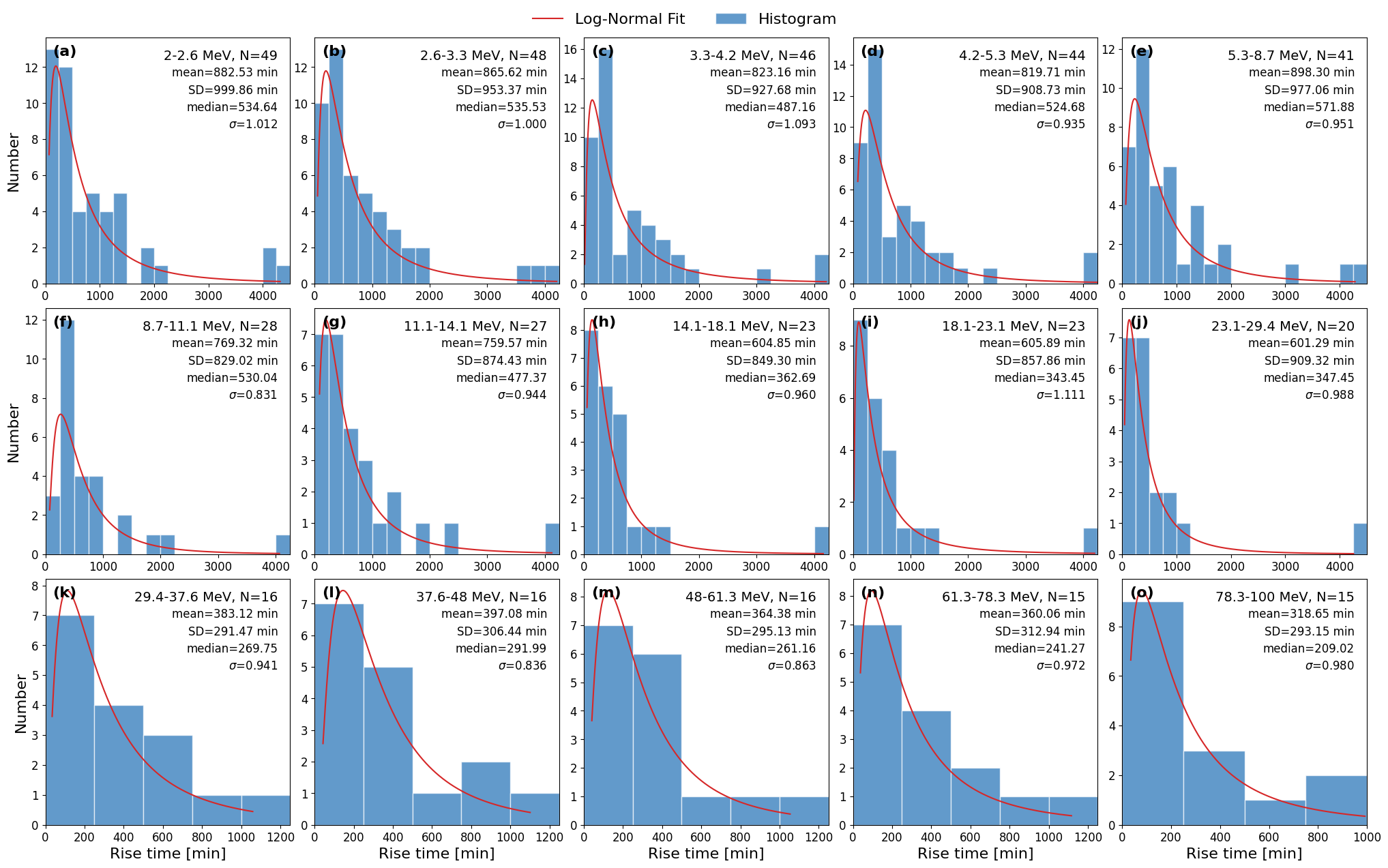} 
\caption{Histograms of the rise times of SEP events observed by the Tianwen-1/MEPA instrument at the orbit of Mars across different energy channels. Panels (a)-(o) correspond to 15 energy intervals spanning 2–100 MeV. In each panel, the blue histogram (bin width is fixed at $250$ minutes) represents the distribution of rise times, while the red curve shows the log-normal fitting of the distribution. The legend in each panel lists the number of events analyzed $(N)$, the average (mean) and standard deviation (SD) of the observed rise times, and the fitted log-normal parameters with median ($e^\mu$) and shape parameter ($\sigma$) defined in Eqn. \ref{eqn3: log-normal function}.}
\label{fig:2}
\end{minipage}
\end{figure*}

\begin{figure*}[ht!]
\begin{minipage}[t]{1.0\linewidth}
\centering
\includegraphics[width=1.0\hsize]{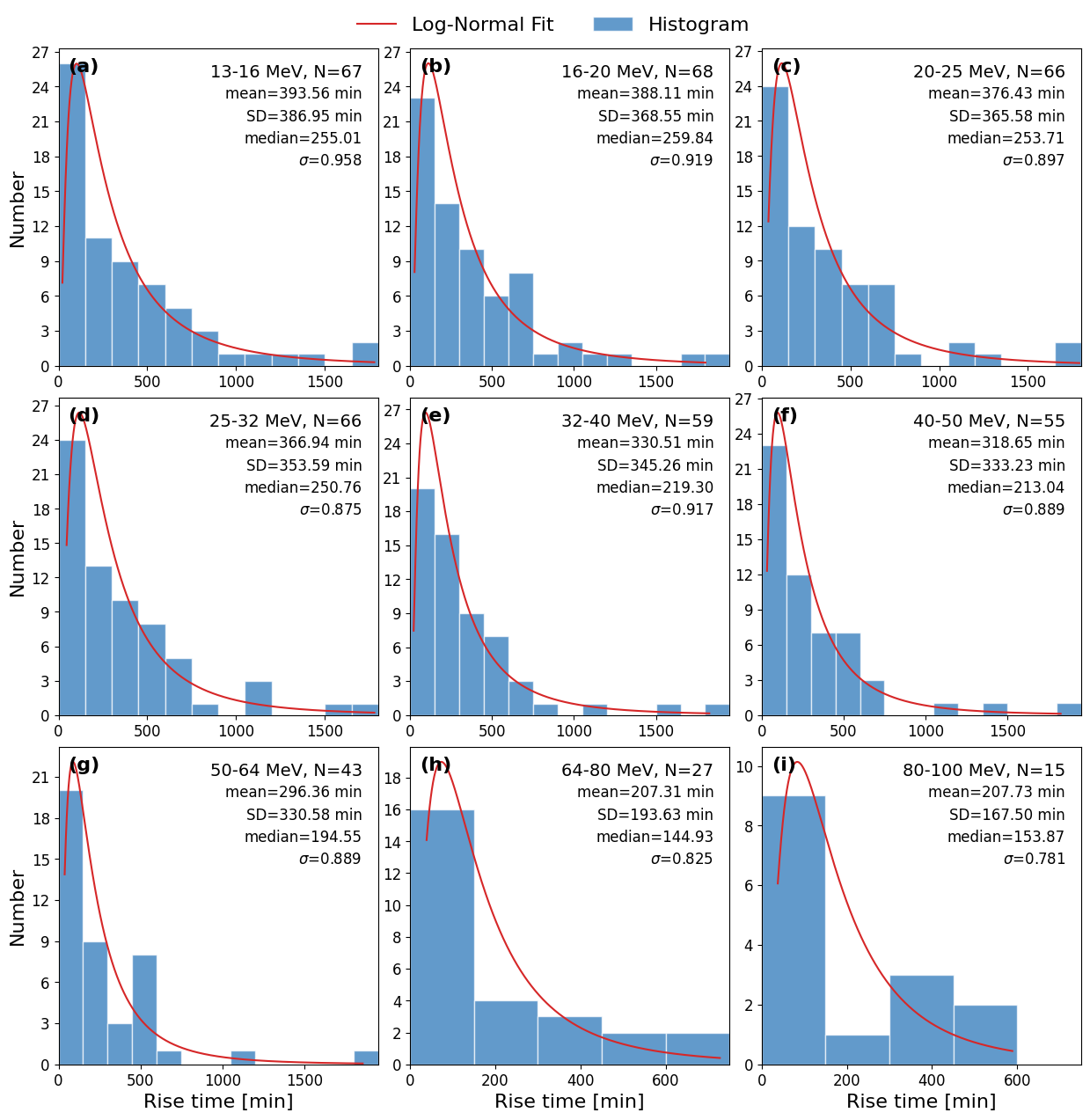} 
\caption{Histograms of the rise times of SEPs observed by SOHO/ERNE across different energy channels. Panels (a)-(i) correspond to 9 energy intervals spanning 13–100 MeV with $200$ minutes of bin width. The parameters shown in legends are the same as in Fig.\ref{fig:2}.}
\label{fig:3}
\end{minipage}
\end{figure*}

\subsection{Statistical analysis of SEP rise Time}
Applying the methodology described above, we systematically extracted the rise time ($\Delta t$) for each energy channel and for each SEP event. Figures \ref{fig:2} and \ref{fig:3} present the statistical distribution histograms of rise times across different energy channels observed by Tianwen-1/MEPA and SOHO/ERNE, respectively. 
It is visible in Fig. \ref{fig:2}(b-f) that there is a sharp increase of the distribution at small rise-time values followed by a decline towards larger rise-time values. This feature is not visible in other panels (at higher energies) since all histograms plotted here share the same bin width of 250 min to maintain the consistency of the following analysis. But the feature becomes visible once the bin-width is set smaller (not shown here) in other panels of Figs. \ref{fig:2} and \ref{fig:3}. 
To quantitatively describe the right-skewed shape of these distributions, a log-normal distribution was fitted to each histogram. The probability density function is given by:
\begin{align}
    f(x)=\frac{1}{x\sigma\sqrt{2\pi}}\text{exp}\left(-\frac{(\text{ln}x-\mu)^2}{2\sigma^2}\right), \label{eqn3: log-normal function}
\end{align}
where $\sigma$ controls the dispersion and $e^\mu$ corresponds to the geometric mean which equals to median in a log-normal distribution. In the legends of Figs.\ref{fig:2} and \ref{fig:3}, parameter $e^\mu$ is denoted as median in the legends. Both the arithmetic mean and median values are then used for the power‑law fits presented in Fig.\ref{fig:4}.

The statistical results reveal the following facts: (1) There is a significant negative correlation between the rise time and the particle energy, i.e., higher energies correspond to shorter rise times. For example, the average rise time of SEPs at Mars decreases from $882.53$ minutes at the lowest energy (2 MeV) to 318.65 minutes at the highest energy (100 MeV). A similar trend is observed near Earth, with the rise time decreasing from 393.56 minutes at 13 MeV to 207.73 minutes at 100 MeV. (2) Comparing approximately the same energy intervals, the rise times near Mars are generally longer than those near Earth. This increase is consistent not only with the longer transport path at greater heliocentric distances but also with the radial evolution of scattering conditions inferred from our analysis. These observational trends and their inferred dependencies point to underlying physical processes that will be examined in Sect.\ref{sec:results}.

\section{Theoretical model based on observational results}\label{sec:results}
\subsection{SEP rise time derived from the pure diffusion model} \label{sec:results.1}
A relatively complete description of the transport and modulation of SEPs in the expanding solar wind is described by the Fokker-Planck equation \citep{parker1965passage} as follows:
\begin{align}
   \frac{\partial f}{\partial t} = -\frac{V_{sw}}{r^2} \frac{\partial}{\partial r}\left(r^2 f\right) + \frac{2V_{sw}}{3r} \frac{\partial}{\partial E}\left[f E n(E) \right] + \frac{1}{r^2} \frac{\partial}{\partial r}\left(\kappa r^2 \frac{\partial f}{\partial r}\right), \label{eqn4.1: transport_function}
\end{align}
where $f(r, E, t)$ represents the distribution of the particles in space, energy, and time; $V_{sw}$ is the solar wind speed; $\kappa$ is the diffusion coefficient; $m_0$ is the rest mass of the particle; and $n(E)$ represents the normalization conversion factor between particle momentum and kinetic energy, defined as $n(E)=p \, dE / (E \, dp)$. For relativistic particles, $n(E)\approx1$; and for non-relativistic particles, $n(E)\approx 2$. The three terms on the right-hand side represent the key physical processes that govern SEP transport: convection, adiabatic cooling, and diffusion.  Each of these transport mechanisms may modify the original acceleration characteristics of energetic particles.

\citet{wang2022quantitative} simplified this equation under the assumptions of instantaneous impulsive injection and pure radial diffusion. Assuming the radial diffusion coefficient follows a power-law form $\kappa=\kappa_0(r/r_0)^\beta$ with $0\leq \beta<2$, they obtained the solution:
\begin{align}
    f(t)=a(t-t_0)^{-c}\text{exp}\left(-\frac{1}{b(t-t_0)}\right), \label{eqn4.1: transport_solution_wang}
\end{align}
where the parameter $b$ and the dimensionless parameter $c$ are given by:
\begin{align}
    b=\frac{(2-\beta)^2\kappa}{r^2}, \quad c=\frac{3}{2-\beta}. \label{eqn4.1: parameter_b_c}
\end{align}

The rise time from onset to peak was then derived as:
\begin{align}
    \Delta t = t_p - t_0=\frac{1}{bc}. \label{eqn4.1: rising_time}
\end{align}

Further analysis by \citet{wang2022quantitative} revealed an empirical power-law relation between the parameters $b$ and $c$ from observational fits $c\propto b^{-\gamma}$, with an observed value of $\gamma\approx 0.4$ and a theoretical expectation of $\gamma=0.5$. Based on their framework, we combine Eqn.\ref{eqn4.1: rising_time} with the relation $c\propto b^{-\gamma}$ to derive the dependence of the rise time on the diffusion coefficient:
\begin{align}
    \Delta t\propto b^{-(1-\gamma)}\propto \kappa^{-(1-\gamma)}. \label{eqn4.1: relationship_dt_k}
\end{align}
This equation establishes the inverse power-law relationship between the rise time and the diffusion coefficient, which will be combined with the energy dependence of $\kappa$ to derive the $\Delta t-E$ scaling in the following subsection.

\subsection{The relationship between rise time and particle energy} \label{sec:results.2}
According to quasi-linear theory (QLT) developed by \citet{jokipii1966cosmic}, the perpendicular diffusion coefficient $\kappa_{\perp}$ is usually much smaller than the parallel one $\kappa_{\parallel}$ \citep{giacalone1999transport}. Numerical simulations further quantify this disparity, showing that the ratio $\kappa_{\perp}/\kappa_{\parallel}$ ranges from $10^{-1}$ to as low as $10^{-4}$ depending on turbulence properties of the interplanetary medium \citep{conlon1978interplanetary, droge2010anisotropic, qin2012numerical}. A comprehensive review of the analytical theories for perpendicular transport, including quasi-linear theory, non-linear guiding center theory, and unified non-linear transport theory, has been provided by \citet{shalchi2020perpendicular}. Furthermore, as discussed by \citet{van2021turbulent}, the drift and perpendicular diffusion mechanisms governing perpendicular transport of SEPs are two primary processes that operate on top of parallel transport, and drift effects become apparent only after $>2$ hours after the event onset. Therefore, during the rise phase, radial diffusion is dominated by parallel diffusion along the magnetic field. Following \citet{jokipii1971propagation}, the radial diffusion coefficient can be expressed as:
\begin{align}
\kappa_r=\kappa_{\parallel}\cos^2(\psi)+\kappa_{\perp}\sin^2(\psi), \label{eqn4.2: kappa_radial}
\end{align}
where $\psi$ is the angle between the radius vector from the Sun and the outward direction along the magnetic field given by $\psi=\text{tan}^{-1}(r\Omega_s/V_{sw})$ (with $\Omega_s=2.9\times10^{-6}\ \text{sec}^{-1}$ being rotation frequency of the Sun and $V_{sw}$ is the solar wind speed). Given that $\kappa_\perp \ll \kappa_\parallel$, we approximate $\kappa_r \approx \kappa_\parallel \cos^2(\psi)$. Since the $\cos^2(\psi)$ factor is purely geometric and does not introduce additional energy dependence, we also absorb this geometric factor into the proportionality constant and use $\kappa\approx\kappa_\parallel$ in the following derivation of the energy dependence.

In particle transport theory, the parallel diffusion coefficient can be expressed as \citep{jokipii1971propagation, droge2010anisotropic}:
\begin{align}
    \kappa_\parallel=\frac{1}{3}v\lambda_\parallel, \label{eqn4.2: kappa}
\end{align}
where $v$ is the particle speed and $\lambda_\parallel$ is the parallel mean free path. 

To establish the relationship between $\Delta t$ and Energy $E$, we examine the energy dependence of $v$ and $\lambda_\parallel$.
Considering that the rest energy of a proton is $E_0=938$ MeV and its kinetic energy is $E$, the total energy is $E_{tot}=E+E_0$. Defining a dimensionless quantity $x=E/E_0$, its speed in the relativistic regime is:
\begin{align}
    v=c_s\frac{\sqrt{x^2+2x}}{1+x}, \label{eqn4.2: velocity}
\end{align}
where $c_s$ denotes the speed of light.

\citet{bieber1994proton} noted that the mean free path follows a power-law dependence on particle rigidity $R$:
\begin{align}
    \lambda_\parallel\propto R^\alpha, \label{eqn4.2: lambda}
\end{align}
where the index $\alpha$ is linked to the background magnetic turbulence spectrum. Theoretically, the turbulence spectrum follows $P(k) \propto k^{-q}$ and $\alpha = 2-q$. Classical Kolmogorov turbulence ($q=5/3$) corresponds to $\alpha=1/3$, while the Iroshnikov-Kraichnan spectrum ($q=3/2$) for magnetohydrodynamic turbulence gives $\alpha=1/2$ \citep{jokipii1971propagation, qin2006effect, wang2015simulations}. 
Two extreme cases are the Bohm diffusion limit (maximum scattering, $\alpha=1$; see \citet{hussein2014detailed} for a numerical investigation in strong turbulence) and rigidity-independence scattering ($\alpha\approx0$) \citep{bieber1994proton, palmer1982transport, chen2024parallel}. Based on ion observational data, \citet{droege2000rigidity} reported a typical $\alpha\approx 0.3$ in the rigidity range of about $30-300$ MV (corresponding to proton kinetic energies of approximately $0.5-47$ MeV). Therefore, $\alpha$ typically ranges between 0 and 1, depending on the state of heliospheric turbulence.

Considering proton rigidity as $R=pc_s=E_0\sqrt{x^2+2x}$, Eqn.\ref{eqn4.2: velocity} becomes:
\begin{align}
    \lambda_\parallel \propto \left(\sqrt{x^2+2x}\right)^\alpha. \label{eqn4.2: lambda2}
\end{align}

Combining the expressions for speed and mean free path gives:
\begin{align}
    \kappa\propto v\lambda_\parallel\propto\frac{\left(x^2+2x\right)^{\frac{1+\alpha}{2}}}{1+x}. \label{eqn4.2: lambda3}
\end{align}

For protons, $E_0=938$~MeV, so for normal SEPs with energies below about 100 MeV, we can assume that $E\ll E_0$ or $x\ll 1$, which leads to $\sqrt{x^2+2x}\approx\sqrt{2x}$ and $1+x\approx 1$. The above equation becomes: 
\begin{align}
    \kappa \propto (2x)^{\frac{1+\alpha}{2}}\propto E^{\frac{1+\alpha}{2}} \label{eqn4.2: kappa2}
\end{align}

Substituting this into Eqn.\ref{eqn4.1: relationship_dt_k} (the empirical correlation between  $\Delta t$ and $\kappa$) yields the final theoretical expression linking rise time $\Delta t$ to particle energy and the turbulence parameter $\alpha$:
\begin{align}
    \Delta t \propto E^{-\frac{(1-\gamma)(1+\alpha)}{2}}. \label{eqn4.2: kappa3}
\end{align}
We further define the power-law exponent $\eta$ as $\Delta t\propto E^\eta$, with $\eta=-(1-\gamma)(1+\alpha)/2$. This formulation makes it evident that, for typical values of $\gamma$ and $\alpha$, the rise time decreases with increasing particle energy. An empirical estimate of $\eta$ derived from observational data is presented below.

\subsection{Comparative analysis of SEP rise time at Earth and Mars} \label{sec:results.3}

\begin{figure}[ht!]
\centering
\includegraphics[width=1.0\hsize]{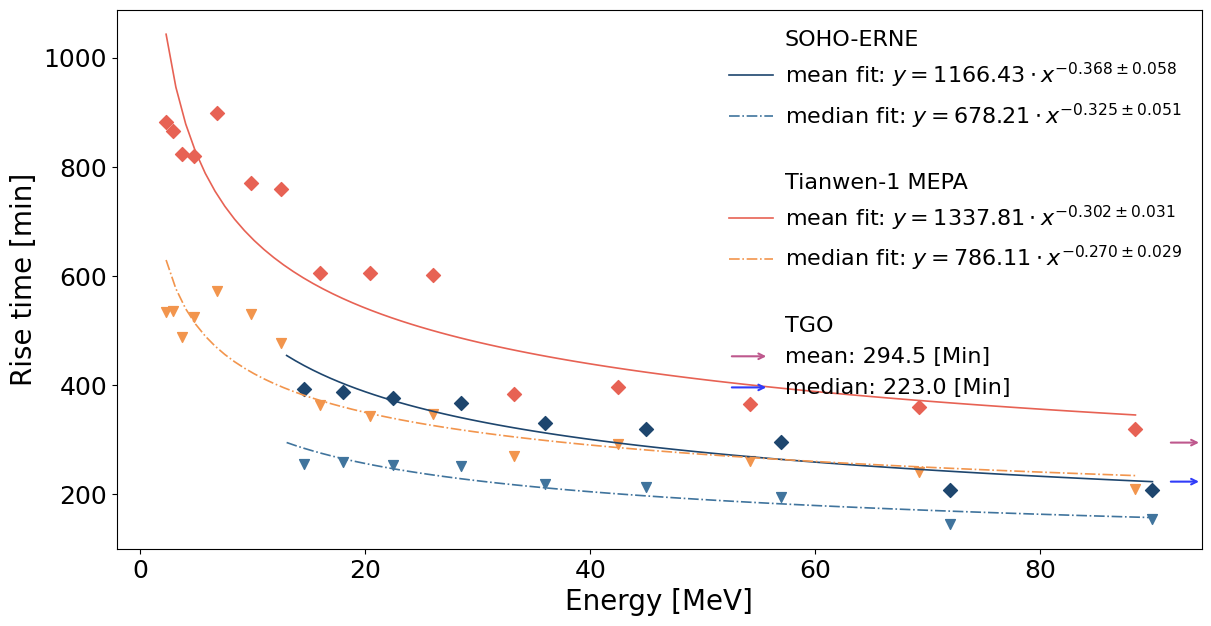} 
\caption{Relation between rise time $\Delta t$ (y-axis) and particle energy $E$ (x-axis) based on statistical observations from Earth and Mars. Diamonds and inverted‑triangle symbols represent the statistical mean and geometric mean from log‑normal distribution fits, respectively. Red and yellow denote MEPA data, dark green and light blue denote ERNE data. The purple and dark blue arrows indicate the mean and median rise time derived from TGO dose-rate measurements.}
\label{fig:4}
\end{figure}

\begin{figure*}[ht!]
\begin{minipage}[t]{1.0\linewidth}
\centering
\includegraphics[width=1.0\hsize]{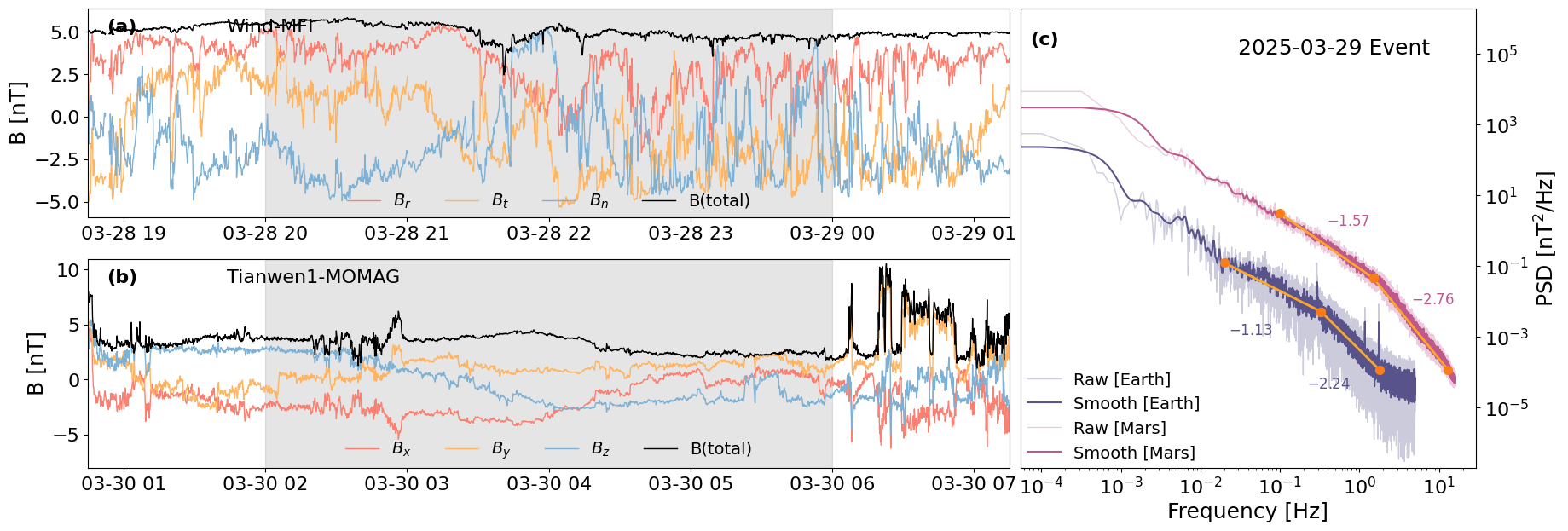} 
\caption{Comparison of PSD of magnetic field at Earth and Mars during the SEP event on March 28, 2025. Panels (a) and (b) show the three components of the magnetic field and the total field magnitude observed by Wind (near Earth) and Tianwen-1 (near Mars), respectively. The gray area marks the 4-hour interval (that encompass the onset time of the SEP event) of magnetic field data used for the PSD analysis shown in panel (c), where the blue and red lines represent the PSD of magnetic field fluctuations at Earth and Mars, respectively. The orange lines indicate the fitted slopes in the inertial and dissipation ranges, with the corresponding spectral indices labeled nearby.}
\label{fig:5}
\end{minipage}
\end{figure*}

Figure \ref{fig:4} shows the rise time as a function of proton energy based on multi-spacecraft observations. The SOHO/ERNE (dark blue lines) and Tianwen-1/MEPA (red and yellow lines) data points represent measurements of different energy, while the TGO dose-rate measurements (arrows) provide independent constraints on SEP rise time at higher energies. The TGO results are consistent with those from the Tianwen-1 MEPA data at high energies, providing a good confirmation of the reliability of our rise time analysis. However, since the dose rate records particles of all energies reaching the detector and does not have a clearly-defined effective energy range, TGO results are not included in the power-law fit plotted as lines in the figure. 

The power-law fit follows the form $\Delta t\propto E^{\eta}$, consistent with the analytical solution of Eqn.\ref{eqn4.2: kappa3}. We fit $\eta=-0.368\pm 0.058$ near Earth and $\eta=-0.302\pm 0.031$ near Mars based on the mean rise time, where the uncertainties represent the $1\sigma$ statistical errors derived from the fitting procedure. Based on the median rise time, we have $\eta=-0.325\pm 0.051$ near Earth and $\eta=-0.270\pm 0.029$ near Mars. However, note that due to different energy limits of different instruments, the fitted energy ranges differ between Earth (13$\sim$100 MeV) and Mars (2$\sim$100 MeV). Harmonizing the energy coverage and accounting for cross‑instrument differences in future work would help better quantify the associated uncertainty in $\eta$. Despite of the uncertainties, the power-law exponent $\eta$ is steeper at Earth than at Mars based on both mean and median results, indicating a stronger energy dependence of the rise time at 1 AU.  

Based on the power-law relations between rise time and energy obtained from the fits in Fig.\ref{fig:4}, we compute the difference of rise times across the 5-100 MeV range at 1 MeV intervals between Mars and Earth. The rise time at Mars is, on average, approximately 137.13 minutes longer than that at Earth. Furthermore, the data and fitted functions reveal that the rate at which rise time decreases with increasing energy gradually flattens at higher energies. This behavior can be understood in terms of energy losses during radial propagation. As a direct manifestation of particle transport, the rise time is sensitive to both particle speed and its evolution with distance. \citet{cao2025radial} demonstrated that the particle flux decreases with distance as $f(R)\propto R^{-\phi(E)}$, where the decay index $\phi(E)$ declines with increasing energy. Consequently, lower energy particles lose more energy during transport through processes such as diffusion and adiabatic cooling, resulting in a greater reduction in speed and thus a prolonged rise time, while higher energy particles suffer less energy loss and exhibit shorter rise times. At sufficiently high energies, energy losses become negligible compared to the total energy of particles, and the rise time is primarily governed by velocity, exhibiting only a weak dependence on energy.

While energy loss considerations offer a qualitative understanding of the observed relationship between rise time and energy, we provide a quantitative interpretation within the framework of a parallel diffusion model. Adopting the theoretical relation $\eta=-(1-\gamma)(1+\alpha)/2$ derived in Eqn.\ref{eqn4.2: kappa3} and assuming $\gamma=0.5$ as theoretically expected considering primarily the diffusion process \citep{wang2022quantitative}, we invert the observed power-law exponents to obtain the turbulence parameter $\alpha$. The uncertainty in $\alpha$ can be propagated from the fitting error of $\eta$ via:
\begin{align}
    \sigma_\alpha = \left|\frac{\partial \alpha}{\partial \eta}\right|\sigma_\eta = \frac{2}{1-\gamma}\sigma_\eta \simeq 4\sigma_\eta. \label{eqn4.2: alpha_uncertainty}
\end{align}
Using the statistical uncertainty of $\eta$ from the mean-value fits, we obtain $\alpha=0.472\pm 0.232$ at Earth and $\alpha=0.208\pm 0.124$ at Mars. The geometric-mean fits yield very similar results: $\alpha=0.3\pm 0.204$ at Earth and $\alpha=0.08\pm 0.116$ at Mars, respectively. The difference between $\alpha$ derived from observations at Earth and Mars indicates a considerable weakening of the rigidity dependence of the parallel mean free path with increasing radial distance. The smaller $\alpha$ value at Mars suggests that turbulence in that region is more similar to a rigidity-independent scattering regime than that near Earth.

This radial evolution of $\alpha$ reflects fundamental changes in the properties of solar wind turbulence. As the distance increases, the break frequency separating the inertial range from the dissipation range shifts toward lower frequencies in both fast and slow solar wind \citep{bruno2014radial, duan2018angular, duan2020radial, lotz2023radial}. Consequently, at a fixed particle gyro-frequency, the sampled turbulence spectral index $q$ becomes larger, leading to a reduction in $\alpha$ via $\alpha=2-q$.

This radial evolution can be assessed by comparing the turbulence spectra observed during an SEP event on March 28, 2025. In this event, Earth and Mars were nearly aligned along the same magnetic field line, providing a unique opportunity to isolate the radial evolution of turbulence. Figure \ref{fig:5} presents the PSD of magnetic field fluctuations at Earth (Wind-MFI) and Mars (Tianwen1-MOMAG). For statistical reliability, we analyze 4-hour intervals of magnetic field data that encompass the onset time of the SEP event at both locations. As shown in panel (c), the spectral indices at Earth are -1.13 in the lower frequency range (likely corresponding to the inertial range) and -2.24 in the higher frequency range (approaching the dissipation range). At Mars, the spectrum is systematically steeper, with indices of -1.57 and -2.76 in the lower and higher frequency ranges, respectively. This feature is consistent with the smaller $\alpha$ value derived from the rise‑time obtained for Mars in Sect.\ref{sec:results.3}. However, the break frequency at Mars is higher than that at Earth during this event, which differs from the aforementioned radial evolution trend based on solar wind observations. 
This discrepancy can be explained by energy injection in the Martian upstream solar wind. As noted by \citet{marino2023scaling}, such injection (possibly driven by pickup ions or foreshock waves) can induce inverse cascades, which enhance dissipation at smaller scales (higher frequencies) and thus shift the break frequency toward higher values.

Beyond this single event, the diversity of turbulence spectra observed near Mars provides a broader statistical context for the radial trend inferred from our SEP analysis of rise time. \citet{zou2025statistical} classified magnetic PSD upstream of Mars into three types (Types A, B, and C) based on spectral morphology. Type A spectra exhibit a well-developed inertial range followed by a dissipation range, resembling those commonly observed near Earth. Energy cascades through the inertial range and dissipates via ion-scale mechanisms, such as cyclotron damping \citep{chen2014ion}. Type B spectra show evidence of energy injection at some scales, resulting in a steeper dissipation range. Type C spectra lack a distinct dissipation range, suggesting that dissipation may occur at frequencies beyond the observational window. Notably, only about one-quarter of events at Mars display Type A spectra similar to those near Earth, while over one-third fall into Type B and another third into Type C \citep{zou2025statistical}. The radial shift of the dissipation range is further supported by \citet{cheng2022variation}, who found that the Taylor scale, which marks the transition from the inertial range to the dissipation range, is approximately one order of magnitude smaller in the upstream solar wind of Mars than at Earth. 
These studies suggest a substantial fraction of solar wind magnetic field exhibits intrinsically steeper spectra at Mars due to break frequency shift and/or different turbulence regimes (energy injection or absence of dissipation range). Therefore, the smaller $\alpha$ we derive at Mars is consistent with a steeper effective turbulence spectrum, whether resulting from a transition into the dissipation range or from fundamentally different spectral morphologies.

Several additional factors may further contribute to the observed radial trend. The weakening of the interplanetary magnetic field with distance, together with possible changes in its anisotropic structure, alters the resonance matching conditions between waves and particles and thereby weakens the rigidity dependence of the scattering efficiency. These processes jointly lead to particle transport behavior that approaches a ''rigidity-independent” regime.

In addition to the power-law analysis presented above, the rise time data from TGO and MEPA shown in Fig.\ref{fig:4} can be used to examine the characteristic energy at which the radiation dose and particle flux exhibit the strongest correlation in Mars orbit. The TGO AB dose measurements yield a mean rise time of $294.5$ min and a median of $223.0$ min as shown in the figure. Comparing these values with the power-law curves obtained from fitting the MEPA data (the mean fit $y=1337.81E^{-0.302\pm0.031}$ and the geometric mean fit $y=786.11E^{-0.270\pm0.029}$) gives intersection energies of $150.15$ MeV and $106.31$ MeV, respectively. Propagating the uncertainty in the power-law exponent $\eta$ yields energy ranges of $[94.1, 266.3]$ MeV for the mean and $[67.6, 186.2]$ MeV for the median. These ranges can be considered as the main SEP proton energy range contributing to the radiation dose measured by TGO orbiting Mars. 
Meanwhile, the TGO data provide an important high energy constraint on the relationship between rise time and energy.

\section{Discussion and conclusion} \label{sec:conclusion}
This study examined a series of SEP events observed by the SOHO/ERNE (near Earth) and Tianwen‑1/MEPA (near Mars) instruments between November 2020 and March 2025. We selected 75 events in Earth orbit and 58 events in Mars orbit that satisfied our identification criteria. For each event, we used the linear fitting method to determine the onset time and an optimized peak-finding algorithm to extract the peak time, and provided statistical estimates of the relationship between rise time and particle energy.

Under the assumptions of instantaneous injection of source particles and pure radial diffusion, where parallel diffusion dominates during the rise phase, we derived from the diffusion model the theoretical relation $\Delta t\propto E^{-(1-\gamma)(1+\alpha)/2}$. 
Meanwhile, observations show that the rise time follows a distinct power‑law dependence on energy: we derive $\Delta t\propto E^{-0.368\pm0.058}$ near Earth and $\Delta t\propto E^{-0.302\pm 0.031}$ near Mars. Inverting the turbulence parameter $\alpha$ through $\eta=-(1-\gamma)(1+\alpha)/2$, we infer $\alpha=0.472\pm 0.232$ at Earth and $\alpha=0.208\pm0.124$ at Mars. The smaller $\alpha$ value at Mars indicates that the mean free path is less dependent on rigidity with increasing distance. This trend aligns with turbulence evolution: while turbulence near Earth is predominantly in the inertial range, at larger distances (about 1.5 AU) it shifts toward a dissipation range dominated, more rigidity independent scattering regime.

The present model relies on several simplifying assumptions that could introduce uncertainties into the inverted parameters. A key assumption is instantaneous particle injection, which does not capture the complexity of acceleration at CME-driven shocks. In diffusive shock acceleration (DSA), particles gain energy gradually through repeated interactions with the shock front, with higher-energy particles requiring longer acceleration times \citep{drury1983introduction, blandford1987particle}. This energy-dependent acceleration time naturally leads to a scenario where higher energy particles are released later than lower energy ones, which may be the reason causing the recent observations of inverse velocity arrival (IVA) detected by Parker Solar Probe and Solar Orbiter \citep[e.g.,][]{li2025delayed, chen2025evidence}. 
On the other hand, CME-driven shocks can extend particle acceleration over prolonged periods, leading to continuous injection of SEPs \citep[e.g.,][]{reames1999particle, desai2016large}. As the shock propagates outward into the interplanetary space, its strength gradually decreases reducing the maximum energy attainable by accelerated particles while continuously  accelerating particles to lower energies. 
Together, these effects tend to prolong the observed rise times, more likely at lower energies, leading to a steeper inferred energy dependence ($\eta$ in $\Delta t\propto E^\eta$) and a potential overestimation of the turbulence parameter $\alpha$ compared to the instantaneous-injection scenario.

Second, the diffusion model employed in this study considers mainly parallel diffusion along the mean magnetic field. However, both parallel and perpendicular diffusion operate during SEP transport in reality, and the degree of magnetic connectivity varies from event to event. When the spacecraft is poorly connected to the acceleration point, perpendicular diffusion across field lines could become an essential transport process. Neglecting this mechanism may bias the inferred transport parameters, particularly for poorly connected events. Furthermore, the apparent rise time reflects not only local scattering conditions but, more importantly, the time required for particles to propagate across field lines, in particular during poorly connected events, adding further complexity to its interpretation.

Third, the pure diffusion model employed in this study does not explicitly include adiabatic cooling. While adiabatic cooling does not directly affect the rise time itself, it could play a crucial role in shaping the particle energy spectrum during propagation \citep{wang2024statistical}. As demonstrated by \citet{cao2025radial}, adiabatic cooling modulates the rate at which particle flux decreases with distance, leading to a radial dependence of spectral hardening that varies with the source spectrum. This energy dependent modulation may alter the relative arrival times of particles at different energies, thereby influencing the power-law exponent $\eta$ and, consequently, the inferred turbulence parameter $\alpha$. In Sect.\ref{sec:results.3}, we invoked this effect to qualitatively account for the observed reduction in $\eta$. However, a more complete assessment would require a transport model that simultaneously accounts for diffusion, adiabatic cooling, and their interplay.

Last but not least, the selection of events and the peak-detection algorithm, though objective in procedure, retain some subjectivity in parameter choices, which could somehow bias the rise time statistics. Nevertheless, the use of rise time rather than absolute flux reduces the impact of cross‑calibration uncertainties and provides a new perspective in understanding the transport features of SEPs at different heliospheric distances.

Despite the aforementioned uncertainties associated with the model assumptions, the observed power-law relations between rise time and energy at Earth and Mars are statistically significant and represent a reliable observation result. The clear radial evolution of the power-law exponent $\eta$ and the derived turbulence parameter $\alpha$ suggests that the underlying physical trend, namely the weakening of the rigidity dependence with radial distance, is unlikely to be an artifact of model simplifications. Nevertheless, as the interpretation is based on a diffusion model that neglects certain key physical processes, it is subject to quantitative uncertainties. More refined models incorporating finite injection of particle source, perpendicular diffusion, and adiabatic cooling will be essential to disentangle the contributions of source and transport processes.

Future studies could focus on detailed case‑by‑case analyses of well‑connected SEP events to examine how rise time profiles vary with energy and magnetic connection geometry, thereby providing specific constraints on local turbulence parameters. 
Moreover, combining multi-satellite observations (e.g., Parker Solar Probe, Solar Orbiter, BepiColombo, and outer‑heliosphere missions) would enable the construction of a radial-dependent profile of the turbulence parameter $\alpha$. Such a multi‑point approach would help to disentangle the radial evolution of turbulence from event‑to‑event variations and to verify turbulent transport models of SEPs in the heliosphere which may serve as important tools for space weather predictions.

\begin{acknowledgements}
The authors acknowledge the support by the National Natural Science Foundation of China (Grant Nos. 42130204, 42188101, 42521007, 42474221) and the National Key R\&R Program of China (2025YFF0510900)
The Level-2 archived data from both the MEPA and MOMAG instruments onboard Tianwen-1 are publicly available through the Planetary Exploration Project Science Data Release System at (\url{https://moon.bao.ac.cn/}). In addition, the MOMAG data are also accessible via the official website of MOMAG team at the University of
Science and Technology of China (USTC, \url{http://space.ustc.edu.cn/dreams/tw1_momag/}). 
\end{acknowledgements}

\bibliographystyle{aa}
\bibliography{bib_abbreviated}{}

\begin{appendix}
\end{appendix}

\end{document}